# Luminescence and absorption in short period superlattices

M.F. PEREIRA*

*Department of Condensed Matter Physics, Institute of Physics CAS*
*Na Slovance 1999/2, 182 21 Prague 8, Czech Republic*
*\*email: pereira@fzu.cz*

**Abstract** - This paper applies analytical approximations for the luminescence of short period semiconductor superlattices and analyses the low density regime, demonstrating that the theory clearly connects with low density absorption with ratios of oscillator strengths of bound and continuum states as expected from the Elliott formula. A numerical study illustrates in detail the bleaching of higher order bound state. The analytical expressions have potential for systematic studies of controlled excitonic pathways characterized by THz responses.

## 1 Introduction

A full understanding of how photoexcitations evolve into Coulomb-bound electron and hole pairs, called excitons, and unbound charge carriers is a key cross-cutting issue in photovoltaics and optoelectronics. Furthermore, these excitonic pathways are characterized by THz responses [1]. Consequently, systems where these effects can controlled per design have strong potential for applications. Furthermore, semiconductor materials are the required substrate for Photoconductive Antennas, for which novel efficient solutions are constantly being sought [2] and semiconductor superlattices may become a successful and efficient substrate.

Semiconductor superlattices (SSLs) are media where the effective dimensionality of the electrons and holes can be controlled between two and three dimensions and are thus very important media to investigate transport and optical effects [3,4] from the GHz to the THz-Mid Infrared (TERA-MIR) range [5,6]. The smooth evolution from strongly localized quasi-two dimensional to highly delocalize quasi-three dimensional, notably for the relatively less studied inter-valence band case, may lead to interesting consequences for the coupling with light in nano and microcavities leading to polaritons [7].

The interest on semiconductor superlattices (SSLs) has been renewed due to the possibility Terahertz (THZ) radiation generation per frequency multiplication [8-14] and as model systems for nonlinear dynamics and band transport [15]. If SSL multipliers can be directly integrated with Gunn or Superlattice Electron Devices [16,17], the combination may become an alternative to quantum cascade lasers [18] in the low end of the THz spectrum. For a thorough, recent review and a comparison between electronic and optical sources of THz radiation, see Ref. [19].

It is thus timely to exploit different approaches for the the theoretical description of optical properties of SSLs, such as luminescence, which is a very important tool to characterize new materials. As a matter of fact, recent research has led to analytical solutions of the Dyson equation for the polarization function, which is the selfenergy in the Photon Green's function. Notably, the "s-shape" in the luminescence profiles as a function of temperature for these materials have been studied in ternary GaAs$_{1-x}$Bi$_x$ [20] and InAs$_{1-x}$N$_x$ [21] as well as for more complex quaternary materials, such as InAs$_{1-x-y}$N$_x$Sb$_y$ [22], all in very good agreement with experimental data from different teams [23-26]. In a parallel effort, analytical solutions have shown that absorption nonlinearities increase with the anisotropy in SSLs [27-29].

This paper bridges the gap between these two lines of study and shows luminescence as a function of temperature and increasing anisotropy for short period SSLs with corresponding absorption, shows the connection between the equations and demonstrates that the approach reproduces the Elliott formula for excitons with the correct balance between bound and continuum states in the low density limit.

## 2 Mathematical Formalism

Nonequilibrium Green's Functions (NEGF) techniques including the photon Greens' function lead directly to expressions for the luminescence power spectrum and have reproduced both single beam and pump-probe luminescence experiments accurately [30-32]. They also lead to generalized semiconductor Bloch equations, can be applied to both intersubband [33-35] and interband transitions [36], but in both cases, require intensive numerical methods. The usual scattering mechanisms that affect the carriers transport are described by selfenergies [33] which lead to the absorption and luminescence linewidth. However, as highlighted in the introduction, analytical solutions based on controlled approximations have been recently developed. It is beyond the scope of this paper to repeat extensive derivations delivered in the references cited above. The important information to highlight is that the free photon Green's function represents the photons propagating without any interaction with the medium. When carriers are injected the transverse polarization function P, which is the selfenergy in photon Green's function Dyson equation, determines how the excited medium modifies the photon propagation. The lesser Keldysh component $P^<$ is proportional to the carriers recombination rate and yields the number of emitted photons per unit area. It thus governs the power emission spectrum through the relation

$$I(\omega) = (\hbar\omega^2/4\pi^2 c)\, iP^<(\omega). \tag{1}$$

The imaginary and real parts of $P^r$ are, respectively proportional to absorption/gain and refractive index changes, through the connection with the dielectric function of the medium

$$\epsilon(\omega) = 1 - \frac{c^2}{\omega^2} P^r(\omega). \tag{2}$$

For details, see Refs. [30-32]. Considering that for the excitation conditions of relevance for this paper, the (real) background dielectric constant $\epsilon(\infty) = n_b^2$, is much larger than both real and imaginary parts of the changes induced by the carriers, the absorption coefficient is thus

$$\alpha(\omega) = \frac{c}{2\omega\sqrt{\epsilon(\infty)}} Im\{P^r(\omega)\}. \tag{3}$$

The next step is to combine the Kubo-Martin-Schwinger (KMS) relation under the form derived in Ref. [29]

$$P^<(\omega) = \frac{-2i Im\{P^r(\omega)\}}{1 - \exp[(\hbar\omega - \mu)/(K_B T)]}, \tag{4}$$

can be combined with Eqs. (1), (3) and the solution for the power spectrum [20, 37].

$$I(\omega) = \frac{I_0}{1 + exp(\beta(\hbar\omega - \mu))} \left\{ \sum_{n=1}^{\sqrt{g}} \frac{4\pi}{n} \left( \frac{1}{n^2} - \frac{n^2}{g^2} \right) \delta_\Gamma(\xi - e_n) + 2\pi \int_0^\infty \frac{sinh(\pi g\sqrt{x})}{cosh(\pi g\sqrt{x}) - cos(\sqrt{4g - g^2 x})} \delta_\Gamma(\xi - x) dx \right\}, \tag{5}$$

to deliver the direct connection with the absorption spectrum

$$\alpha(\omega) = \alpha_0 \vartheta \left\{ \sum_{n=1}^{Int\{\sqrt{g}\}} \frac{4\pi}{n} \left(\frac{1}{n^2} - \frac{n^2}{g^2}\right) \delta_\Gamma(\zeta - e_n) + \int_0^\infty \frac{2\pi \, \sinh(\pi g \sqrt{x})}{\cosh(\pi g \sqrt{x}) - \cos(\pi \sqrt{4g - g^2 x})} \delta_\Gamma(\zeta - x) dx \right\}, \qquad (6)$$

where $I_0 = \frac{\hbar \omega^2 e^2 |\langle S|x|X\rangle|^2 (E_g^0/\hbar)^2}{\pi e_0 c^3 a_0^3}$, $e_n = -\frac{1}{n^2}\left(1 - \frac{n^2}{g}\right)^2$, $n = 1,2,3 \ldots$, $\zeta = (\hbar\omega - E_g)/e_0$, $g = (\kappa a_0)^{-1}$ were $\kappa$ is the inverse screening length, and $a_0, e_0$ denote, respectively the exciton Bohr radius and binding energy, $\vartheta = \tanh[\beta(\hbar\omega - \mu)/2]$ and

$$\alpha_0 = \left(\frac{e^2}{\hbar c}\right) \left(\frac{E_g^0}{\hbar\omega}\right)^2 \left(\frac{\hbar\omega}{e_0}\right) \left(\frac{d^2}{a_0^3}\right) \frac{1}{\sqrt{\epsilon(\infty)}} \, . \qquad (7)$$

The first term in the RHS of Eq. (7) is the dimensionless fine structure constant and $e^2 d^2 = e^2 |\langle S|x|X\rangle|^2$ is the dipole moment between the conduction and valence bands at $k = 0$. Note that in the spectral region around the bandgap where excitonic corrections are stronger, $\hbar\omega \sim E_g^0$ and the second term on the RHS of Eq. (7) can be approximated by $\left(\frac{E_g^0}{\hbar\omega}\right)^2 \sim 1$ and consequently, $\alpha_0 \sim \left(\frac{e^2}{\hbar c}\right) \left(\frac{\hbar\omega}{e_0}\right) \left(\frac{d^2}{a_0^3}\right) \frac{1}{\sqrt{\epsilon(\infty)}}$.

The approximation used for renormalized bandgap is [36]

$$E_g = E_g^0 + e_0 \begin{cases} -1 + \left(1 - \frac{1}{g}\right)^2, & g \geq 1 \\ -1/g, & g < 1 \end{cases} \qquad (8)$$

In the low density excitonic limit, $g \to \infty$, $\vartheta \to 1$, $e_n \to -\frac{1}{n^2}$, $E_g \to E_g^0$, leading to the Elliott formula [37]

$$\alpha(\omega) = \alpha_0 \left\{ \sum_{n=1}^\infty \frac{4\pi}{n^3} \delta_\Gamma(\zeta - e_n) + \int_0^\infty \frac{\pi e^{\pi/\sqrt{x}}}{\sinh^{\pi/\sqrt{x}}} \delta_\Gamma(\zeta - x) dx \right\}, \qquad (9)$$

showing that both luminescence and absorption formulas have the correct balance between bound and continuum states.

## 3 Numerical Results

The numerical examples in this section are for short period superlattices with strong delocalization of the electron and hole wavefunctions can be described in many cases by anisotropic 3D media, characterized by in-plane and transverse (along the growth direction) effective masses and dielectric constants. The anisotropy parameter $\gamma$ is given by the ratio between the in-plane $\mu_\parallel$ and perpendicular $\mu_\perp$ reduced effective masses, $\gamma = \mu_\parallel / \mu_\perp$, which are calculated from the non-interacting superlattice Hamiltonian $\mathcal{H}_0$, for electrons and holes, i.e. i = e or h,

$$\frac{1}{\mu_\parallel} = \frac{1}{m_{e\parallel}} + \frac{1}{m_{h\parallel}} \, , \frac{1}{\mu_\parallel} = \frac{1}{m_{e\parallel}} + \frac{1}{m_{h\parallel}} \, ,$$

$$\frac{1}{m_{i\parallel}} = \hbar^{-2} \partial^2/\partial k_{i\parallel}^2 \langle \Psi | \mathcal{H}_0 | \Psi \rangle \, , \quad \frac{1}{m_{i\perp}} = \hbar^{-2} \partial^2/\partial k_{i\perp}^2 \langle \Psi | \mathcal{H}_0 | \Psi \rangle \qquad (10)$$

These can be calculated from the corresponding free carrier Hamiltonian, and full details of technique which delivered good agreement with experimental excitonic data, can be found in Ref. [4].

Figure 1 shows calculated absorption and corresponding luminescence using the parameters determined by anisotropic medium theory for a short period GaAs-Al$_{0.3}$Ga$_{0.7}$As superlattice with repeated barrier and well widths equal to 1.5 nm. The resulting effective masses are $m_{e\|} \approx 0.078$; $m_{h\|} \approx 0.126$, $m_{e\perp} \approx 0.082$ and $m_{h\perp} \approx 0.447$. These lead the anisotropy parameter γ= 0.698. The resulting exciton binding energy and Bohr radius are given respectively by = $e_0 = 5.22$ meV and $a_0 = 11.3$ nm.

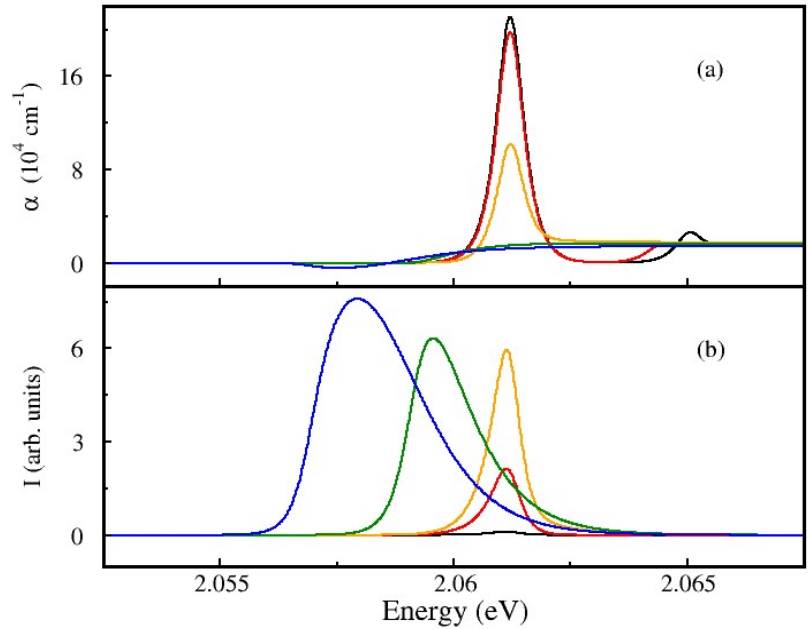

*Fig.1* Absorption and luminescence of a short period GaAs-Al$_{0.3}$Ga$_{0.}$As superlattice with repeated barrier and well widths equal to 1.5 nm. All curves have been calculated with the same broadening $\Gamma = 6.7$ meV and temperature T=10 K. The absorption $\alpha(\omega)$ $of$ $Eq.$ (6) is given in (a) from top to bottom (colour online) black, red, orange, green and blue correspond to carrier densities N=1 × 10$^{13}$, 1 × 10$^{14}$, 1 × 10$^{15}$, 5 × 10$^{15}$ and 1 × 10$^{16}$ carriers/cm$^3$. The luminescence $I(\omega)$ of Eq. (5) is given in (b) using the same colour convention to compare with the corresponding curves in (a).

The correct ratio of oscillator strength between bound and continuum states, highlighted by the exact limiting case in Eq. (9) is clearly seen. Both 1s and 2s states are visible at low density until they bleach out due to many body effects in absorption (Fig1.a), but it is difficult to see them in luminescence. The next figures highlight the effect in more detail.

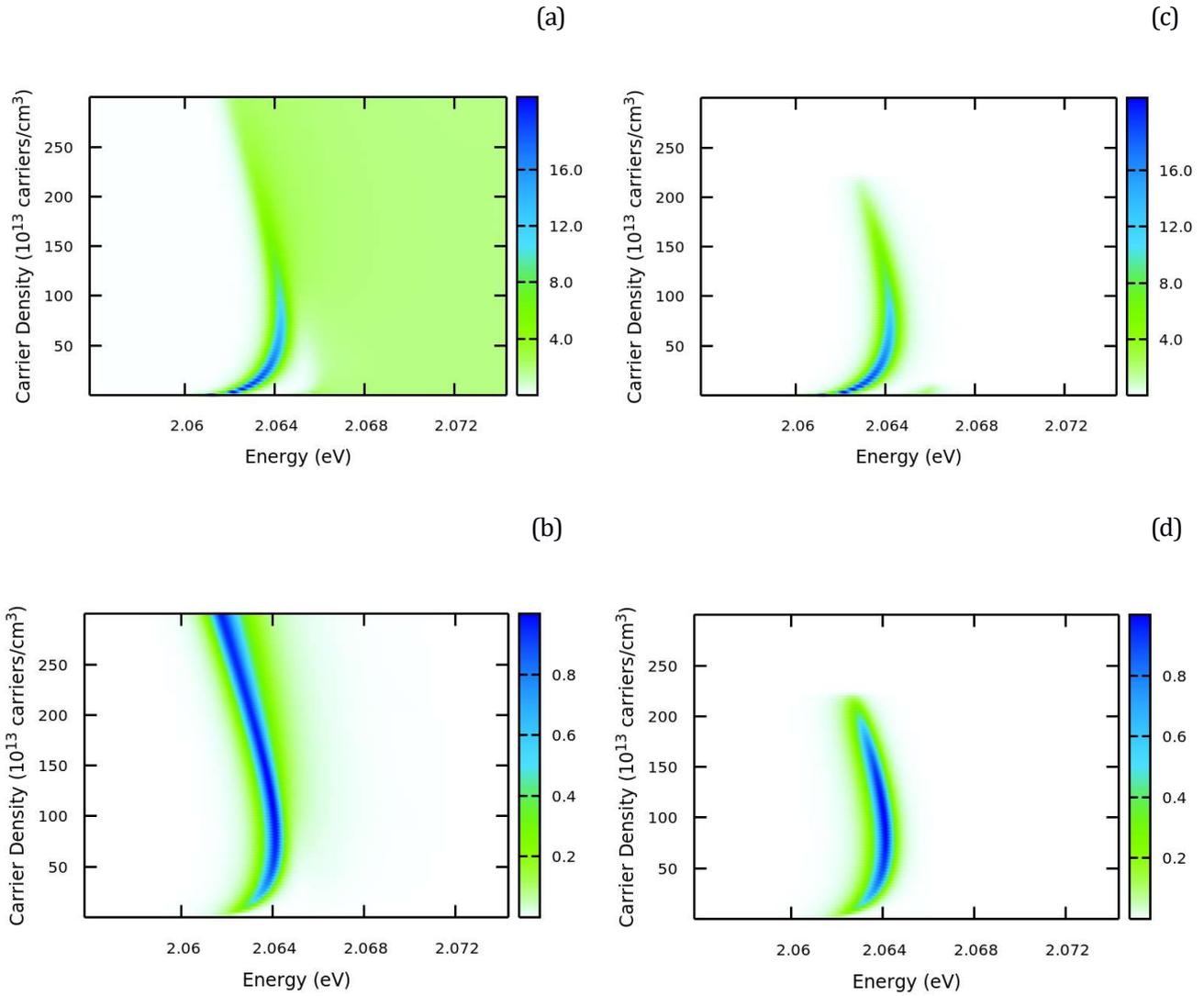

*Fig.2* Absorption $\alpha(\omega)$ from Eq. (6) (a) and (b), compared with the corresponding luminescence from Eq. (5) for the same short period GaAs-Al$_{0.3}$Ga$_0$. As superlattice of Fig. 1 and the same broadening $\Gamma = 6.7$ meV and temperature T=10 K. The full spectrum is shown in (a) and (b) while (c) and (d) have only the bound states contributions. The surface plots for absorption are given in $10^4$ /cm$^{-1}$ units exactly as in Fig. 1.a and the luminescence is in arbitrary units with the maximum scaled to 1.

Note the effect of the continuum in the absorption full green area in Fig.2.a. and the luminesecence at high densities in Fig.2.b. Figures 1.c and 1.d show the contribution of bound states only and they are of course bleached due to many body effects at high densities. Note also the small absorption contributions beyond the 1S state around 2.065 eV in Figures 1.a and 2.c. These do not appear on luminescence on the same scale as the 1S contribution, because luminescence always favours the lowest energy transitions.

Contributions due to absorption beyond the 1S state, are shown in the detail in Fig, 3. The 1S luminescence maximum of Fig. 2.d is approximately 3299 times larger than the maximum of the sum of all bound states luminescence beyond 1S in Fig. 3.b.

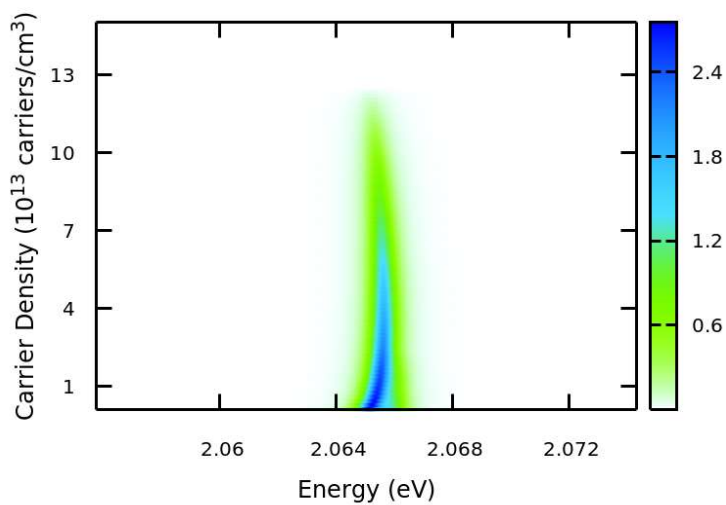

(a)

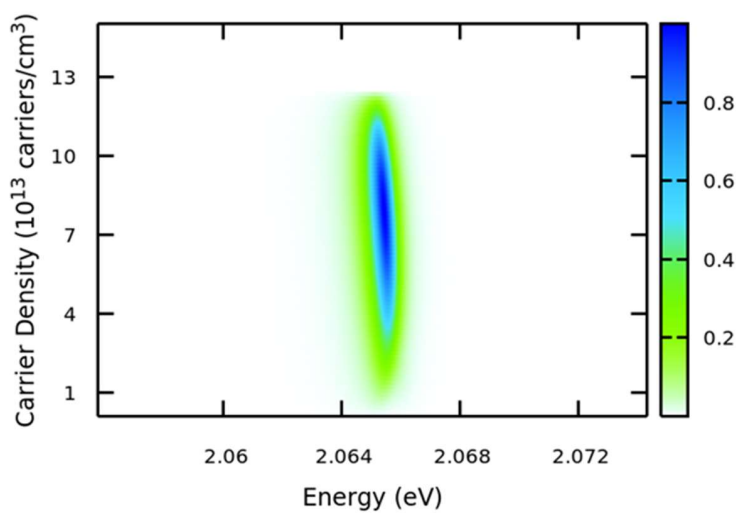

(b)

*Fig.3* Absorption $\alpha(\omega)$ from Eq. (6) (a) and (b), compared with the corresponding luminescence from Eq. (5) for the same short period GaAs-Al$_{0.3}$Ga$_{0.}$ As superlattice of Fig. 1 and the same broadening $\Gamma = 6.7$ meV and temperature T=10 K. Only the bound states contributions beyond the 1S are shown. The surface plot for absorption is given in $10^4$ /cm$^{-1}$ units exactly as in Fig. 1.a and the luminescence is in arbitrary units with the maximum scaled to 1. The 1S luminescence maximum of Fig. 2.d is approximately 3299 times larger than the maximum of the sum of all bound states luminescence beyond 1S in Fig. 3.b.

Before summarizing the results obtained in this paper, two important points should be noted: an excellent review on many-body correlations and excitonic effects in semiconductor spectroscopy is given in Ref. [40], which has very useful relations and a connection to Elliott's formula, but to the best of our knowledge, the expressions imtroduced in Refs. [20, 37] and extended here, are the only fully explicit analytical solutions for the luminesce with Coulomb correlations which are directly programmable for bulk and superlattices in the anisotropic medium limit. Furthermore, effects such as the Urbach tail are only simulated by the choice of linewidth function $\delta_\Gamma$. A predictive description is possible by including interactions with phonons and localized states by momentum independent selfenergies, such as those used in Ref. 10. Full frequency and momentum dependent selfenergies would prevent a direct use of the solutions of the Hulthén potential used in this paper.

In summary, this paper bridged the gap between luminescence and nonlinear absorption in short period superlattices, in which carrier tunnelling lead to quasi-three dimensional behaviour. The low density limit equations and the numerical results using the full formula demonstrate that the approach reproduces the Elliott formula for excitons with the correct balance. The bandgap renormalization used here is an approximation, but the important feature is to see the luminescence following consistently the evolution from absorption to gain with a simple analytical approach, in contrast to previous intensively numerical calculations. Details of high order bound states and their fast bleaching as a function of injected carrier densities are shown in detail. The accuracy and simplicity of the method should make it a powerful tool for research and development of new materials and devices and can play a role on a systematic control of excitonic pathways characterized by THz responses.

*Acknowledgements: The author acknowledges support from the Programme "IMPA Verão 2018" of the Instituto de Matematica Pura e Aplicada (IMPA) in Rio de Janeiro, Brazil.*